\documentclass[%
 reprint,
 amsmath,amssymb,
 aps,
]{revtex4-1}

\usepackage{graphicx}
\usepackage{dcolumn}
\usepackage{bm}
\usepackage{graphicx,latexsym,color,dsfont,float}
\usepackage{amsmath,amsfonts,amsbsy,amssymb,amsthm}
\usepackage[utf8]{inputenc}
\usepackage[mathscr]{euscript}
\usepackage[T1]{fontenc}
\usepackage{mathtools,physics}
\usepackage[english]{babel}
\usepackage{scalerel}
\usepackage[medium]{titlesec}
\usepackage{bm}
\usepackage[bookmarksnumbered=true,bookmarksopen=true]{hyperref}
\hypersetup{
	colorlinks=true,
	linkcolor=blue,
	filecolor=blue,      
	urlcolor=blue,
	citecolor=blue
}

\begin{document}


\title{Gravitational Collider Physics via Pulsar-Black Hole Binaries}

\author{Qianhang Ding$^{1,2}$}
\email{qdingab@connect.ust.hk}
\author{Xi Tong$^{1,2}$}
\email{xtongac@connect.ust.hk}
\author{Yi Wang$^{1,2}$}
\email{phyw@ust.hk}
\affiliation{${}^1$Department of Physics, The Hong Kong University of Science and Technology, \\
	Clear Water Bay, Kowloon, Hong Kong, P.R.China}
\affiliation{${}^2$The HKUST Jockey Club Institute for Advanced Study, The Hong Kong University of Science and Technology, \\
	Clear Water Bay, Kowloon, Hong Kong, P.R.China}

\begin{abstract}
We propose to use pulsar-black hole binaries as a probe of gravitational collider physics. Induced by the gravitation of the pulsar, the atomic transitions of the boson cloud around the black hole back-react on the orbital motion. This leads to the deviation of binary period decrease from that predicted by general relativity, which can be directly probed by the Rømer delay of pulsar time-of-arrivals. The sensitivity and accuracy of this approach is estimated for two typical atomic transitions. It is shown that once the transitions happen within the observable window, the pulsar-timing accuracy is almost always sufficient to capture the resonance phenomenon.
\end{abstract}

\maketitle


\section{Introduction}
The study of elementary particles usually involves the observation of collision events between accelerated particles. In the traditional high-energy physics, this is accomplished by colliders of ever-increasing size and scale \cite{CEPCStudyGroup:2018rmc,Abada:2019lih,Djouadi:2007ik,Linssen:2012hp}. However, its limitations from the energy frontier and the weak-coupling frontier motivate us to find possible alternatives elsewhere on the vast atlas of modern physics. In particular, gravity itself is capable of doing the heavy-lifting. Namely, when the particle mass $\mu$ is close to the background Riemann curvature scale, $\mu^2\sim R$, Schwinger-like effect tends to produce particles, with detectable properties encoded in their later evolution. In the past a few years, indeed, various exciting ideas aiming at testing particle physics in a background gravitational field have been proposed and developed \cite{Chen:2009zp,Baumann:2011nk,Arkani-Hamed:2015bza,Baumann:2018vus,Baumann:2019ztm}. For instance, the cosmological collider operates during inflation with $R\sim H^2\sim (10^{22-23}\text{GeV})^2$, and it serves as a possible extension of the energy frontier \cite{Chen:2009zp,Baumann:2011nk,Arkani-Hamed:2015bza}. 

More recently, Gravitational Collider Physics (GCP) proposed in \cite{Baumann:2019ztm} operates near a Black Hole (BH) with $R\sim (GM_B)^{-2}\sim (10^{-11}\text{eV})^2$ (for a solar-mass BH), where $M_B$ is the BH mass. Since the interaction is mediated by gravity, it serves as a possible extension of the weak-coupling frontier and a tool to probe ultralight bosons such as axions. The basic underlying idea is as follows. Around a spinning BH, superradiance instability occurs when the mass of a boson is smaller than the curvature scale, $i.e.$, $\alpha\equiv G M_B\mu<1$. Then certain energy levels around the Newtonian potential of the BH is abundantly populated via the instability, and this combination of BH and bosonic cloud is often referred to as a gravitational atom. For an isolated BH, the bosonic cloud simply grows via superradiance and depletes via Gravitational Wave (GW) emission, without much observational effects. However, when coupled to a binary companion, the gravitational mixing effect can trigger atomic transitions between different energy levels, in the form of Landau-Zener resonances. The backreaction to the binary system produces floating or sinking orbit which can be observed using GW detectors such as LIGO, LISA and PTA as a distinctive frequency/period dependence $f_{\text{GW}}(t)=2f(t)=2P(t)^{-1}$.

While GCP is initially proposed for binary BHs and GW detectors, we point out that this is not the only observation channel. As the gravitational coupling is universal, the same resonances should still happen if the binary companion of the BH is replaced by any other astronomical object with similar mass. In particular, the binary companion can be a pulsar (PSR). Moreover, the resonance phenomenon is, by itself, irrelevant to the emitted GW, therefore, we are not restricted to observing GWs only. It is known that the timing accuracy of pulsar binaries can be high enough for the \textit{indirect} detection of inspiral-phase GWs \cite{Hulse:1974eb}, long before LIGO directly observed the first merger event \cite{Abbott:2016blz}. As a result, observing a PSR-BH binary through ordinary electromagnetic wave channel serves as an alternative \textit{direct} probe of GCP.

Although there has not yet been any observation of PSR-BH binaries, and the estimated number of such binaries within our Galaxy is limited \cite{Faucher_Gigu_re_2011, Shao:2018qpt}, we argue that as long as there is a clear detection of PSR-BH pair with suitable mass ratio and orbital period, the timing of the orbital phase using Rømer delay is almost always accurate enough to probe the GCP resonance. With the rapid technological advances in radio astronomy in the recent years \cite{wang2014xinjiang,dewdney2009square,Arzoumanian:2020vkk}, we expect this PSR-BH-radio channel to be an important channel for probing gravitational collider physics.

Note that there are existing studies on the physics of ultralight bosons with PSR-BH binaries. For example, \cite{Kavic:2019cgk} aims at detecting outspirals caused by the fast energy loss of boson clouds via GW radiation, while \cite{Seymour:2020yle} focuses on the change in the orbital period derivative due to an extra Yukawa-like force mediated by the boson. We stress our proposal is based on the resonance backreaction mechanism of GCP and is of a different origin from those in the literature. In reality, all three effects may be present and need to be thoroughly investigated. We leave such extensive studies to the future work.

This paper is organized as follows. First, we give a lightning review of GCP in Sect.~\ref{GCPrev}. In Sect.~\ref{SandA}, we examine the sensitivity of radio telescopes to pulsars and compare them with that of the GW detectors. We also study the timing accuracy of PSR-BH binaries. Then in Sect.~\ref{GCPsection}, we estimate the detectability of GCP resonances via the PSR-BH-radio channel. We conclude in Sect.~\ref{conclusion}. Throughout this paper, we use natural units and set $\hbar=c=1$.

\section{A brief review of GCP}\label{GCPrev}
In this section, following \cite{Baumann:2019ztm}, we briefly review the basic elements of gravitational collider physics and collect the useful ingredients for later usage. The starting point is an isolated Kerr black hole and a light boson field. For simplicity, throughout the paper, we treat the ultralight boson as a scalar field. For low-energy bound states near the BH, the scalar field is effectively non-relativistic and satisfies the Schr\"{o}dinger equation
\begin{equation}
	i\partial_t\psi(t,\vec{x})=\left(-\frac{1}{2\mu}\partial_{\vec{x}}^2+V(r)\right)\psi(t,\vec{x})~,
\end{equation}
where $V(r)=-\frac{\alpha}{r}+\mathcal{O}(\alpha^2)$ is the Newtonian potential with relativistic corrections and $\alpha\equiv GM_B\mu$ is the gravitational fine structure constant. Solving the above equation using in-going boundary condition at the BH horizon gives complex-frequency modes, indicating superradiance instability that can lead to the formation of a boson cloud around the BH. More specifically, these modes are labeled by their principle quantum number $n$, angular momentum quantum number $l$ and their azimuthal quantum number $m$ ($e.g.$, $|nlm\rangle$). We are concerned with two of the leading growing modes, $|211\rangle$ and $|322\rangle$. These two energy levels grows and depletes over a time scale
\begin{eqnarray}
	\nonumber T^{(\text{growth})}_{211}&\sim& \frac{10^6 \text{ yrs}}{\tilde{a}}\left(\frac{M_B}{M_\odot}\right)\left(\frac{0.012}{\alpha}\right)^9\\
	T^{(\text{deplete})}_{211}&\sim& 10^8 \text{ yrs}\left(\frac{M_B}{M_\odot}\right)\left(\frac{0.053}{\alpha}\right)^{15}
\end{eqnarray}
and
\begin{eqnarray}
\nonumber T^{(\text{growth})}_{322}&\sim& \frac{10^6 \text{ yrs}}{\tilde{a}}\left(\frac{M_B}{M_\odot}\right)\left(\frac{0.08}{\alpha}\right)^{13}\\
T^{(\text{deplete})}_{322}&\sim& 10^8 \text{ yrs}\left(\frac{M_B}{M_\odot}\right)\left(\frac{0.18}{\alpha}\right)^{20}~.
\end{eqnarray}
Here $\tilde{a}\lesssim 1$ is the dimensionless BH spin.

Now we can introduce a binary companion to exert periodic gravitational perturbation on the scalar cloud. The perturbation can be decomposed into different multipole components, each capable of inducing resonances in the cloud that take the form of Landau-Zener transitions. For quasi-circular equatorial orbits, energy conservation and selection rule gives the orbital period during a Bohr transition $a\to b$ as
\begin{equation}
	P_r=2\pi\left|\frac{\Delta m_{ab}}{\Delta E_{ab}}\right|,~\Delta E_{ab}=\frac{\mu \alpha^2}{2}\left(\frac{1}{n_a^2}-\frac{1}{n_b^2}\right)~,\label{ResonanceP}
\end{equation}
where $\Delta m_{ab}=m_b-m_a$ is the change in azimuthal quantum number. The width of the resonance is
\begin{equation}
	\frac{\Delta P_r}{P_r}=2R_{ab}\frac{q}{1+q},~R_{ab}\lesssim 0.3~.
\end{equation}
During the resonance, the impact on the orbit modifies the changing rate of the orbital period by a factor,
\begin{equation}
	\dot{P}=-\frac{96}{5}(2\pi)^{8/3}(G M_{B})^{5/3}\frac{q}{(1+q)^{1/3}}P^{-5/3}\times\frac{1}{1\pm \delta}~,\label{resPdot}
\end{equation}
where
\begin{equation}
\delta=\frac{\Delta t_c}{\Delta t}\simeq\frac{1}{4}\frac{(1+q)^{4/3}}{q^2}\left(\frac{\alpha}{0.07}\right)\frac{S_{c,0}}{M_B^2}~
\end{equation}
represents the backreaction. Here $S_{c,0}$ is the initial angular momentum of the cloud (see Table 1 in \cite{Baumann:2019ztm}). $\Delta t$ is the original resonance duration and $\Delta t_c$ is the extra resonance duration caused by backreaction. Positive sign in (\ref{resPdot}) gives floating orbit while negative sign gives sinking orbit. Thus the total resonance duration is $\Delta t\pm \Delta t_c$. 

The physical picture is also quite simple. The atomic transitions are accompanied by energy exchanges between the cloud and the binary companion. For $\Delta E_{ab}<0$, the cloud releases its energy so as to cause the binary companion to sink more slowly, hence the name floating orbit. On the other hand, if $\Delta E_{ab}>0$, the cloud drains energy from the binary companion and the companion sinks faster toward the center, and thereby follows a sinking orbit. These floating/sinking orbits possess distinctive deviations from the orbital period derivative predicted from General Relativity (GR). This is the characteristic feature of GCP. Furthermore, observing multiple resonances successively can even help to coin down the detailed evolution history and distinguish bosons with different spin.

\section{Sensitivity and accuracy}\label{SandA}
\subsection{Radio sensitivity}
In order to examine the detailed PSR-BH binary dynamics, one has to be able to see the pulsar first. The intensity of electromagnetic wave emitted from a pulsar decays with the luminosity distance $d_L$ according to the inverse-squared law \cite{lyne2012pulsar}. Therefore, the radio flux density at the telescope is
\begin{align}
S = \frac{L_\nu}{d_{L}^2}~,
\end{align}
where $L_\nu$ is the pseudoluminosity of the pulsar \cite{BAGCHI_2013}. For our purpose, the observable pulsars are relatively close, $i.e.$, $d_L\lesssim 10^4$ kpc, hence we can treat the spacetime to be flat and static. Typically $S$ is smaller than the background noise, so it is difficult to observe individual pulses. However, noise can be reduced by averaging many pulses over an integrated time and over different frequencies. The minimal flux density after such pulse folding technique is given by \cite{lyne2012pulsar}
\begin{equation}
S_{\text{min}}=\frac{\sigma \beta}{\sqrt{N_p\Delta\nu t_{\text{int}}}}\left(\frac{w}{\tau-w}\right)^{1/2}\frac{T_{\text{sys}}}{G}~.
\end{equation}
For instance, in the case of FAST, $\sigma\sim 9$ is the signal-to-noise threshold ratio, $\beta\sim 1$ is a factor due to digitization loss, $N_p=2$ is the number of polarization channels, $\Delta\nu\sim 700$ MHz is the bandwidth, $T_{\text{sys}}\sim 20$ K is the total noise temperature, $G\sim 26.5\text{ K}/\text{Jy}$ is the antenna gain. Then for $\frac{w}{\tau}\sim 0.05$ and an integration time $t_{\text{int}}\sim 120$ s, the minimal flux density is $S_{\text{min}}\sim 0.0038$ mJy \cite{Smits09}. We require $S>S_{\text{min}}$ such that the pulsar itself within the binary can be detected by a radio telescope with suitable $S_{\text{min}}$.

For comparison, we also consider the GW channel for a PSR-BH binary. The amplitude of the GW signal can be written as \cite{Cai:2017cbj},
\begin{align}
h = \frac{4}{d_{L}} \left(G M_B\right)^{5/3}\frac{q}{(1+q)^{1/3}}\left(\pi f_{GW}\right)^{2/3}~,
\end{align}
where $q=\frac{M_P}{M_B}$ is the mass ratio between the pulsar and the BH. Detection in the GW channel requires its amplitude to exceed the characteristic strain, namely $h>h_{\text{min}}$.

\begin{figure}[h!]
	\centering
	\includegraphics[width=8.5cm]{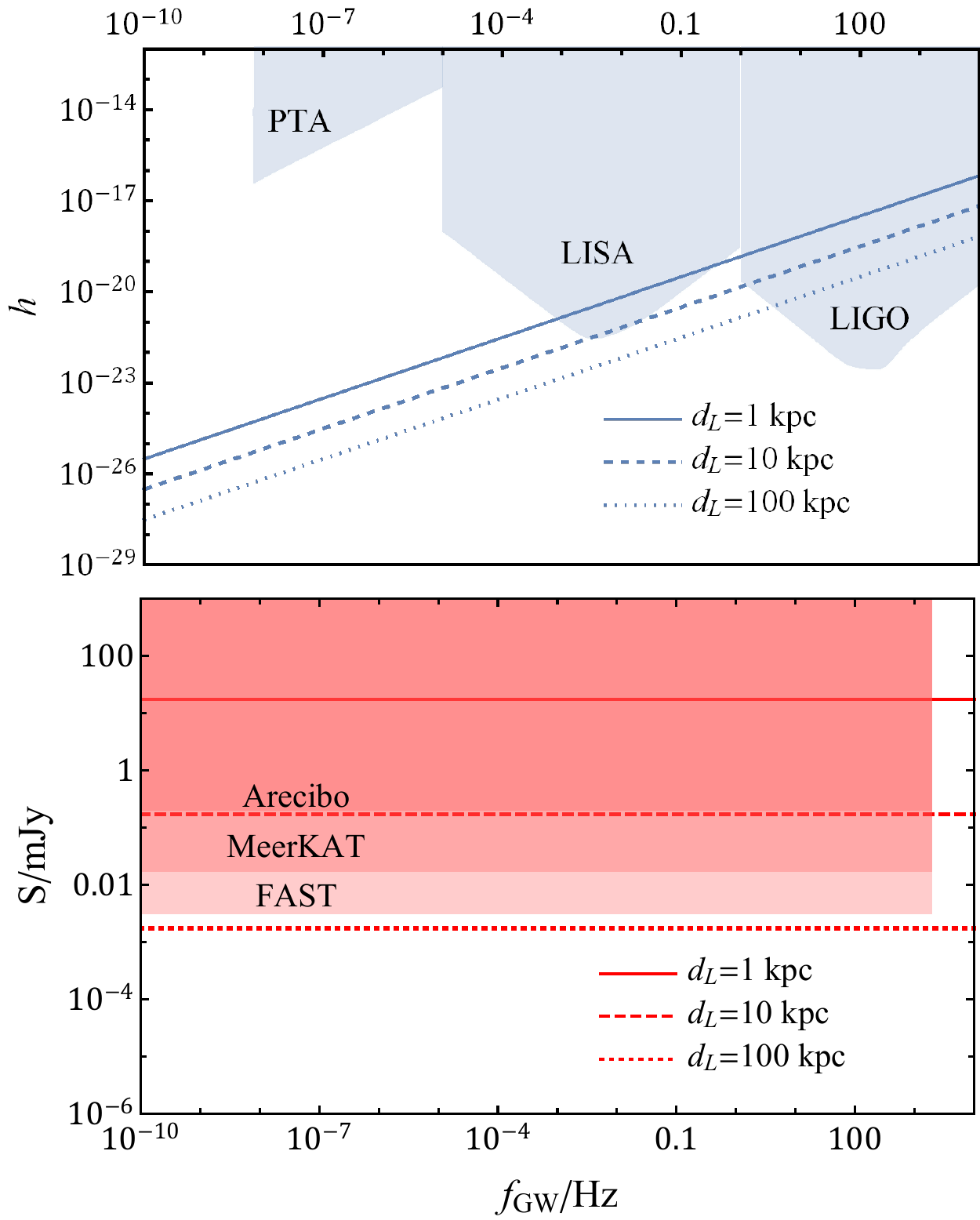}
	\caption{Sensitivity of the GW channel vs sensitivity of the radio channel for a PSR-BH binary. Three different distances typical among the observed pulsars are chosen for comparison. They are plotted using solid, dashed and dotted lines as labeled in the figure. The GW sensitivity curves for LIGO, LISA and PTA are borrowed from \cite{Moore:2014lga}. The minimal radio flux density for Arecibo, MeerKAT and FAST are taken to be 0.2, 0.017 and 0.0038 \cite{lyne2012pulsar, Mauch_2020, Lorimer:2019xjx, Smits09}. Other parameters are chosen as $M_B=1.4M_\odot$, $q=1$ and an average pulsar pseudoluminosity at 1.4 GHz, $L_{1400}=1~\text{mJy kpc}^2$. The high-frequency cutoff of radio telescopes at $f_{GW}=2\times 10^3$ Hz is due to the requirement $1 \text{ ms}\lesssim\tau<P=2f_{GW}^{-1}$ in order that measuring orbital period through Rømer delay is possible. Notice that line segments that lie within the LIGO frequency are thus excluded. But this exclusion is only for solar-mass BHs with $M_B=1.4 M_\odot$. For BHs with different mass parameters, or for resonance happening at different frequencies, the LIGO bound can be evaded. This is the case for GCP resonances that we consider in Sect.~\ref{GCPsection}.}\label{figGWvsPSR}
\end{figure}

We choose a typical set of parameters for the PSR-BH binary and plot the sensitivity of both the GW channel and radio channel in Fig.~\ref{figGWvsPSR}. As is clear from the plot, the radio channel is not much dependent on the binary period, except for the cutoff at pulsar rotation period $\tau$. Its sensitivity is more dependent on pulsar properties such as pseudoluminosity as well as distances. Although the signal-to-noise ratio quickly decreases as $d_L^{-2}$, pulsars with exceptionally large pseudoluminosity can still fall inside the sensitive region. As an example, $d_L=1\text{ Mpc}$ and $L_\nu=10^4\text{ mJy kpc}^2$ gives $S\sim 10^{-2}\text{ mJy}$, which falls within the sensitive region of FAST. For short-distance pulsars ($d_L\lesssim 10^2$ kpc), observations in the radio channel can also be complementary for the blind regions of GW detectors.

\subsection{Timing accuracy}
We now estimate the timing accuracy of PSR-BH binaries. In this subsection, we assume traditional general relativity. The analysis of GCP resonances are left to the next section.

The orbital period is traditionally determined from the periodic Rømer delay of the pulsar. In the computation of Time-of-Arrival (TOA), there are several delays of different origins. Schematically, the orbital delay consists of three parts \cite{Taylor:1994zz},
\begin{equation}
	\Delta_{\text{orb}}\text{TOA}=\Delta_R+\Delta_E+\Delta_S~.
\end{equation}
Here $\Delta_R$ is the Rømer delay that represents the time lapse of light traveling across the binary orbit. $\Delta_E$ is the Einstein delay due to the general-relativistic time dilation within the PSR-BH gravitational field. $\Delta_S$ is the Shapiro delay which is the extra time for light propagation in a curved spacetime. Usually, $\Delta_E$ and $\Delta_S$ are orders of magnitude smaller than the geometrical Rømer delay, but they can still be measured if the TOA precision is high enough. Then the standard procedure to obtain information of the binary is by fitting a given model and minimizing the time residuals using programs such as TEMPO2 \cite{Hobbs:2006cd}. The mass of the pulsar and its companion can be determined from measuring two of five standard Post-Keplerian (PK) parameters, while measuring more than two PK parameters yields a test of general relativity, since the system becomes over-determined \cite{Taylor:1994zz}.

In particular, of the five PK parameters, the orbital period derivative can be directly measured through a recording of the orbital phase shift, or equivalently, the periastron time shift. The periastron time shift is defined as \cite{Suzuki:2019wvg}
\begin{align}
\Delta_{P} (t) = t - P(0) \int_0^{t} \frac{1}{P(t')} dt'~.\label{PerishiftEq}
\end{align}
At a short time scale, $P(t)$ can be approximated by a linear function $P(t) \approx P(0) + \dot{P} t$, with \cite{damour2016gravitational}
\begin{align}
\dot{P}=-\frac{96}{5}(2\pi)^{8/3}(G M_{B})^{5/3}\frac{q}{(1+q)^{1/3}}P(0)^{-5/3}~.\label{GRPdot}
\end{align}
Under the linear approximation, the periastron time shift is simplified to be
\begin{align}
\Delta_{P} (t) = \frac{1}{2} \frac{\dot{P}}{P} t^2~.\label{GRDeltaP}
\end{align}

In order to probe the orbital period change due to GW emission, we need to make sure that the uncertainty of $\Delta_{P} (t)$ measurement is smaller than the cumulative periastron time. 

The uncertainty of periastron time measurement is decided by the single-measurement error and the number of independent measurements. Assuming one can continuously track a pulsar for some time $t_{\text{obs}}\simeq 10$ hrs every day, the single-measurement error comes from phase measurement error in one orbital period and the number of orbital periods within one independent measurement, which can be expressed as $\frac{\tau}{\min (t_{\text{obs}},t)/P}$. We assume the measurement lasts once per day for some time $t<T_{\text{obs}}$, then the number of independent measurements is $\lceil t/1\text{ day}\rceil$, where $T_{\text{obs}}$ is the maximal observation time and $\lceil\rceil$ is the ceil function ($e.g.$, $\lceil 0.1\rceil=1$). The uncertainty of periastron time shift is 
\begin{align}
\sigma_{\Delta_{P}} = \frac{1}{\sqrt{\lceil t/1\text{ day}\rceil}} \frac{\tau}{\min (t_{\text{obs}},t)/P}~.\label{timingAccuracy}
\end{align}

\begin{figure}[h!]
	\centering
	\includegraphics[width=7cm]{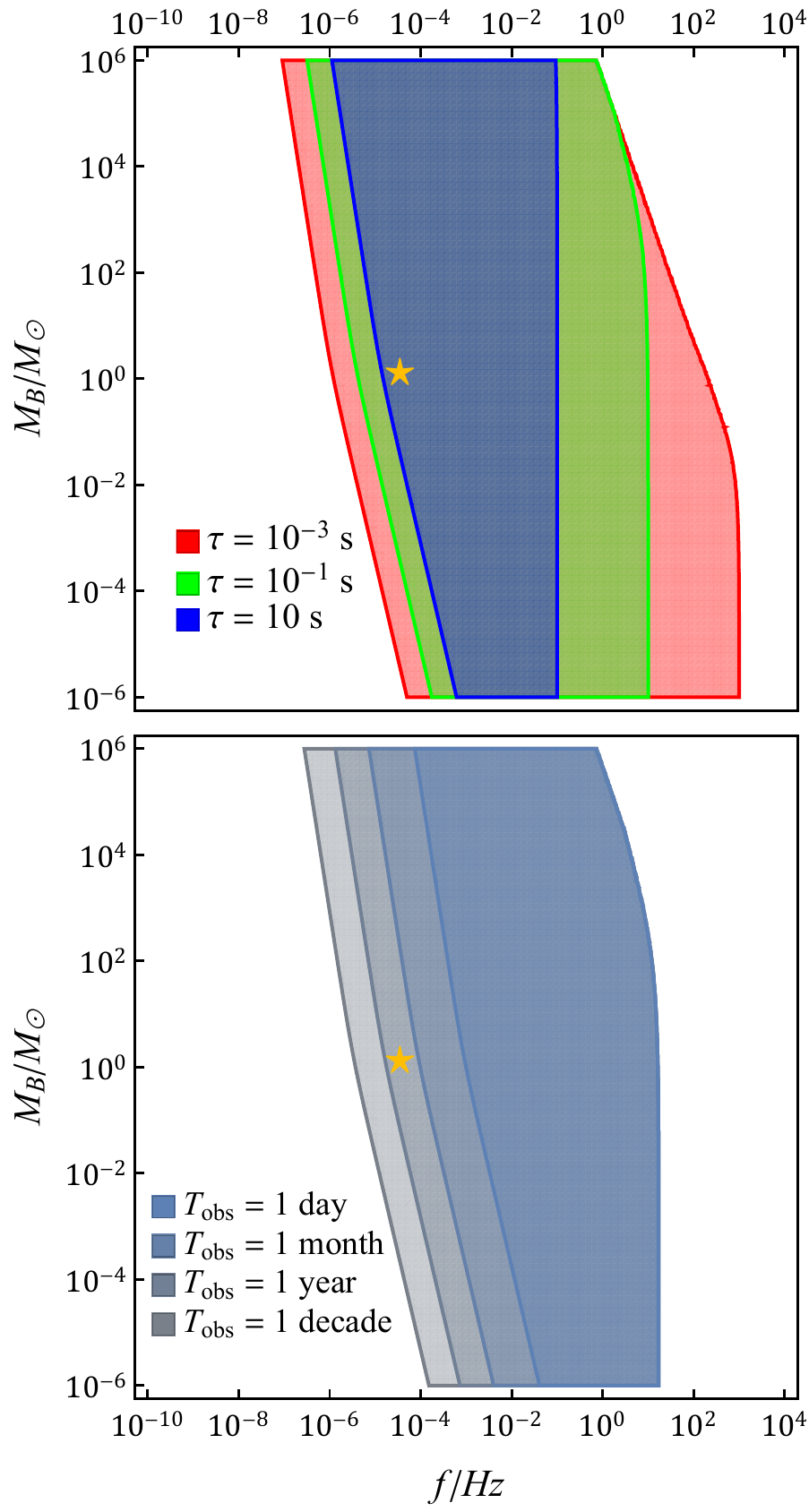}\\
	\caption{Upper panel: Feasible region for pulsars with different rotation periods, with the maximal observation time fixed to be $T_{\text{obs}}=10$ yrs. Lower panel: Feasible region for different maximal observation times, with pulsar rotation periods fixed to be $\tau= 59$ ms. In both plots, the pulsar mass and eccentricity are chosen to be $M_P=1.4 M_\odot$ and $e=0.6$. These parameters match those of the Hulse-Taylor pulsar PSR B1913+16, which is itself labeled by the golden pentagram.}\label{figfMtauAndfigfMTobs}
\end{figure}

To achieve a timing accuracy high enough for probing orbital decay due to GW emission, we must have
\begin{align}
\left|\Delta_{P} (t)\right| >  \sigma_{\Delta_{P}}~.
\end{align}
This crude estimation agrees with the $t^{-5/2}$ scaling behavior for the $\dot{P}$ measurement error \cite{Kramer:2006nb}. In addition, accumulated decrease $|\dot{P} T_{obs}|$ should follow  $|\dot{P} T_{obs}|\ll P$ so that the orbital period decrease is still linear. We also require that the pulsar rotation period is smaller than its orbital period. To observe at least one orbital period, the orbital period cannot exceed the maximal observation time, too. These are the constraints to make our estimation valid. Therefore, we pose the following constraints,
\begin{align}
\nonumber \left|\frac{1}{2}\frac{\dot{P}}{P} t^2\right| &>  \frac{1}{\sqrt{\lceil t/1\text{ day}\rceil}} \frac{\tau}{\min (t_{\text{obs}},t)/P}~,\\
|\dot{P} t |\ll &P~,~  \tau < P< T_{\text{obs}}~.
\end{align}
Notice that the auxiliary variable $t$ should be integrated out. Namely, we take the union of allowed parameter regions for different $0<t<T_{\text{obs}}$.

The resulting estimation of timing accuracy with respect to $f=P^{-1}$ and $M_B$ for different $\tau$ is plotted in Fig.~\ref{figfMtauAndfigfMTobs}~. In addition to astrophysical BHs, here we allow the possibility of Primordial Black Holes (PBHs) with masses smaller than the astrophysical BHs. The left edge of the feasible regions on the $f$-$M_B$ plot represents timing accuracy inadequacy. Therefore choosing pulsars with shorter rotational periods or extending the maximal observation time, as expected, can help to achieve better timing accuracy. The right edge of the feasible regions represents linearity constrains and the validity of our method, namely orbital period must be long enough to leave Rømer delay signatures. In addition, the timing accuracy estimation shows for BH mass lying within $10^{-6}M_\odot<M_B<10^6 M_\odot$, the accessible frequency range is approximately $10^{-7}\text{ Hz}<f<10^3\text{ Hz}$. Notice that this restriction on frequency range cannot be obtained from inspecting radio flux density alone in Fig.~\ref{figGWvsPSR}.

\section{GCP via PSR-BH binaries}\label{GCPsection}
\begin{figure}[h!]
	\centering
	\includegraphics[width=7.5cm]{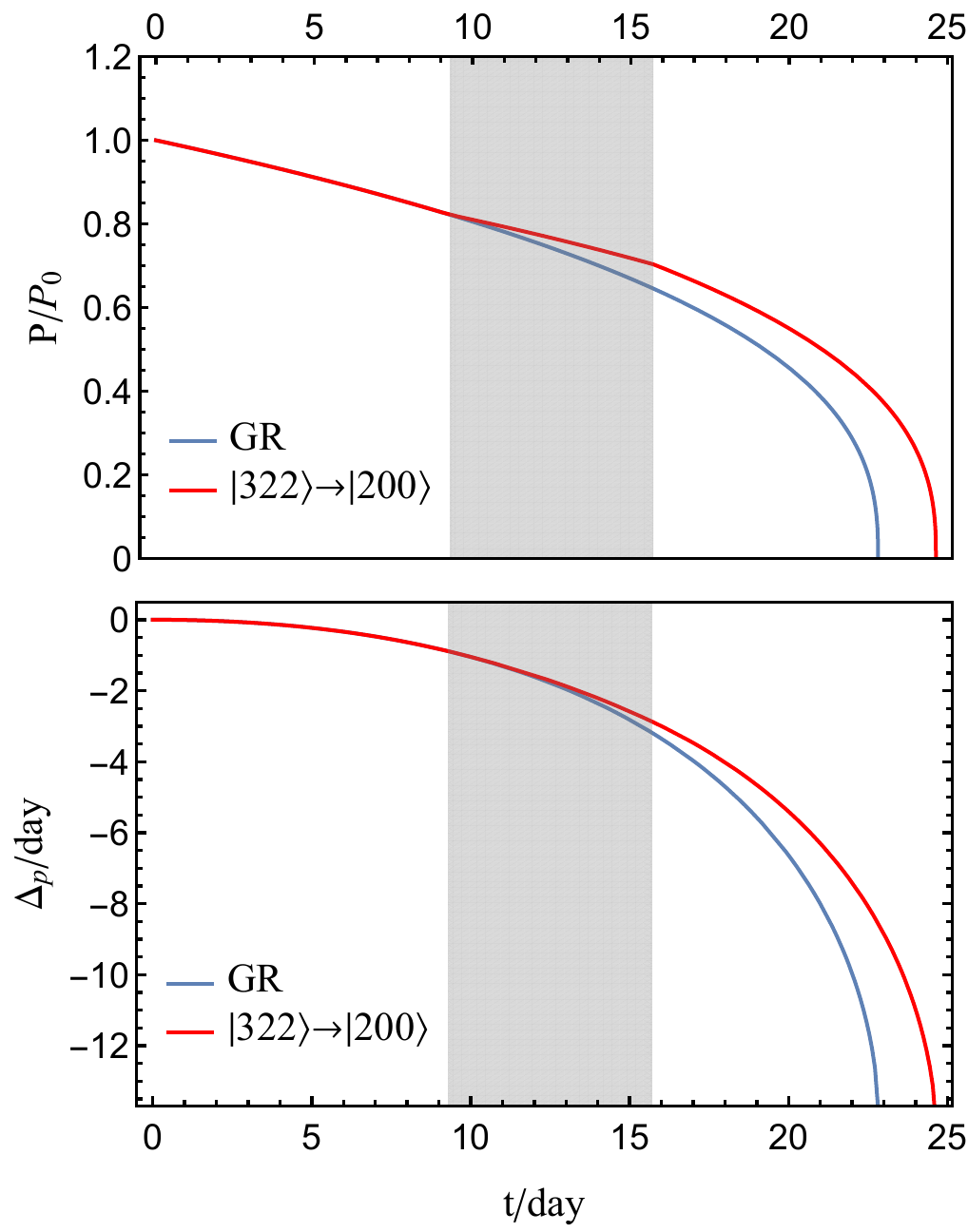}\\
	\caption{Upper panel: Orbital period as a function of time during a floating-orbit resonance $|322\rangle\to |200\rangle$. Lower panel: The deviation in the periastron time shift due to this resonance. The gray region represents the total resonance time $\Delta t+\Delta t_c$. The parameters are chosen as $M_B=4 M_\odot$, $M_P=1.4 M_\odot$, $\alpha=0.1$, $e=0$. The resonance period under this set of parameters is $P_r=3.6$ s.}\label{figPeriastron}
\end{figure}

Now we move on to the detectability of GCP resonances by PSR-BH binaries. The underlying idea is identical to that of the Hulse-Taylor binary pulsar. The only difference here is that the binary period evolution may deviate from that predicted by general relativity alone. In particular, the periastron time shift no-longer follows (\ref{GRPdot}) and (\ref{GRDeltaP}). In the narrow resonance case where different atomic transitions are well-separated, we can approximate the period derivative as
\begin{align}
\nonumber \dot{P}=&-\frac{96}{5}(2\pi)^{8/3}(G M_{B})^{5/3}\frac{q}{(1+q)^{1/3}}P^{-5/3}\\
&\times\frac{1}{1\pm \delta\times\Pi(\frac{P-P_r}{\Delta P_r})}~,\label{GCPevolutionEq}
\end{align}
where $\Pi$ is the Heaviside-Pi window function. To show the qualitative change in the periastron time shift, we solve (\ref{GCPevolutionEq}) for some typical parameter choices and plot the result in Fig.~\ref{figPeriastron}. The detection of periastron time shift evolution of this type can be regarded as a potential GCP signal.

\subsection{Floating orbit: $|322\rangle\to |200\rangle$}
In order to study the allowed parameter space of this PSR-BH-radio channel, we need to pose several constraints. Let us first focus on the floating-orbit resonance $|322\rangle\to |200\rangle$. First, the requirement of fast cloud growth and slow cloud depletion gives \cite{Baumann:2019ztm}
\begin{align}
\text{(C1):  }T^{(\text{growth})}_{322}&\lesssim 10^6 \text{yrs},~~~T^{(\text{deplete})}_{322}\gtrsim 10^8 \text{yrs}~.\label{growthAndDepleteBound}
\end{align}
Second, the resonance orbital period should be larger than the rotation period of the pulsar but smaller than the total observation time, $i.e.$, 
\begin{equation}
\text{(C2):  }	\tau<P_r<T_{\text{obs}}~.\label{Romerbound}
\end{equation}
Third, the Bohr radius $r_0$ of the boson cloud should be smaller than the distance between the pulsar and the BH, in order that the pulsar stays outside the boson cloud,
\begin{equation}
\text{(C3):  }	r_0=\frac{G M_B}{\alpha^2}<R,\text{ with }\frac{R^3}{P_r^2}=\frac{G M_B}{4\pi^2}\frac{q}{1+q}~.
\end{equation}
Combining the above equation with (\ref{ResonanceP}), this third criterion reduces to a simple constraint on the mass ratio, $q>0.0012$. For a pulsar mass $M_P\simeq 1.4 M_\odot$, this gives a bound on the maximal BH mass $M_B\lesssim 1.2\times 10^3 M_\odot$. At last, we require the deviation in the periastron time shift to be larger than the timing accuracy,
\begin{equation}
\text{(C4):  }	|\Delta_P|_{\text{GCP}}-\Delta_P|_{\text{GR}}|>\sigma_{\Delta_{P}}~,\label{GCPtimingbound}
\end{equation}
$\Delta_P|_{\text{GR}}$ and $\Delta_P|_{\text{GCP}}$ are computed using (\ref{GRPdot}) and (\ref{GCPevolutionEq}) respectively. In addition, if $\Delta t+\Delta t_c< T_{\text{obs}}$, it becomes possible to detect a full resonance during our maximal observation time.

\begin{figure}[h!]
	\centering
	\includegraphics[width=7.5cm]{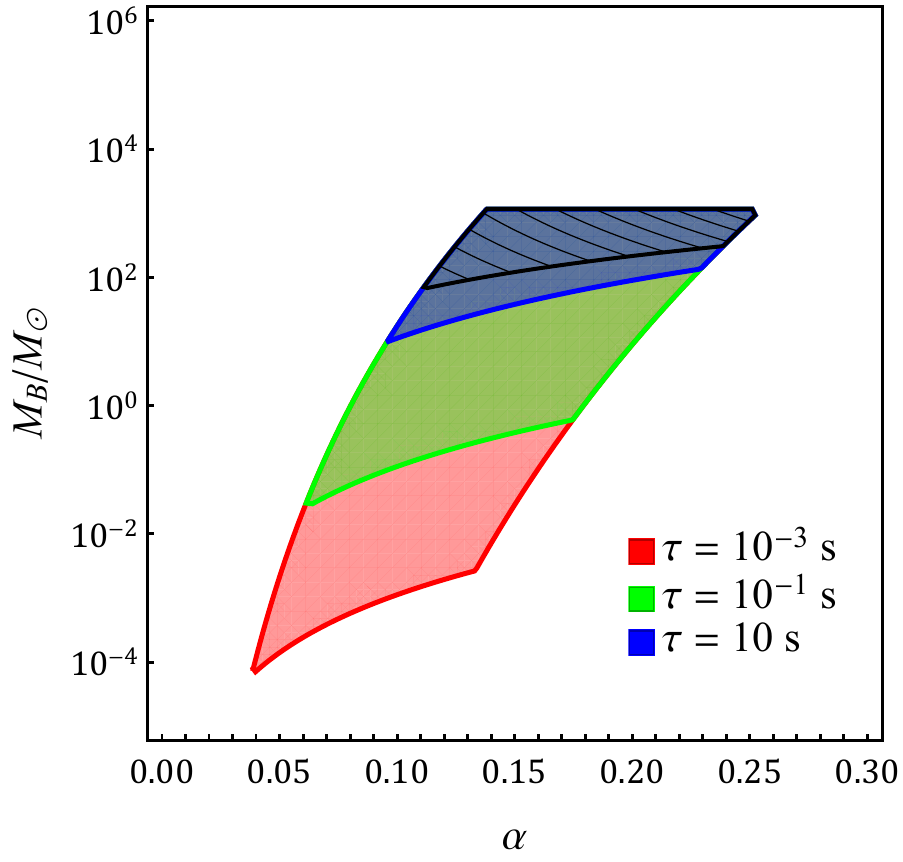}\\
	\caption{The feasible region for $\alpha$, $M_B$ and different pulsar rotation periods $\tau$ in the floating-orbit resonance $|322\rangle\to |200\rangle$. The other parameters are chosen to be $M_P=1.4 M_\odot$ and $T_{\text{obs}}=10 \text{ yrs}$. The shaded area is where $\Delta t+\Delta t_c> T_{\text{obs}}$. Thus we cannot observe a whole resonance for the shaded region but a significant deviation from GR prediction is still detectable.}\label{fig31}
\end{figure}

Combining the above four constraints, we plot the available parameter region in Fig.~\ref{fig31}. Clearly pulsars with shorter rotation periods can reach wider parameter spaces. In fact, the timing accuracy constraint (C4) is looser than the Rømer delay constraint (C2). This means as long as the measurement of orbital period using Rømer delay is possible, the timing accuracy of the PSR-BH system is always enough to capture the extra periastron time shift due to GCP. This is not surprising since for a generic parameter choice in Fig.~\ref{figPeriastron}, the periastron time shift is on the order of days, a significant amount of deviation for clocks so accurate as millisecond pulsars.

We also need to consider the likelihood of detecting a GCP resonance. The absolute probability depends on the number of PSR-BH binaries in the galaxy as well as their distributions of mass, rotation period, luminosity, etc. In \cite{Shao:2018qpt}, a population synthesis calculation shows that the number of PSR-BH binaries in the Galactic disk is around 3-80, of which 10\% may be observed by FAST. However, their estimation rests upon many assumptions. For example, it is assumed that all PSR-BH binaries have an astrophysical origin, rather than for example, scenarios involving PBHs captured by pulsars. Therefore, to reduce these uncertainties and simplify the discussion, we try to examine the \textit{conditional} probability of GCP discovery within $T_{\text{obs}}$, given the fact that a PSR-BH binary with mass $M_B$, $M_P$ and initial period $P_\text{ini}$ has already been observed.

\begin{figure}[h!]
	\centering
	\includegraphics[width=7.5cm]{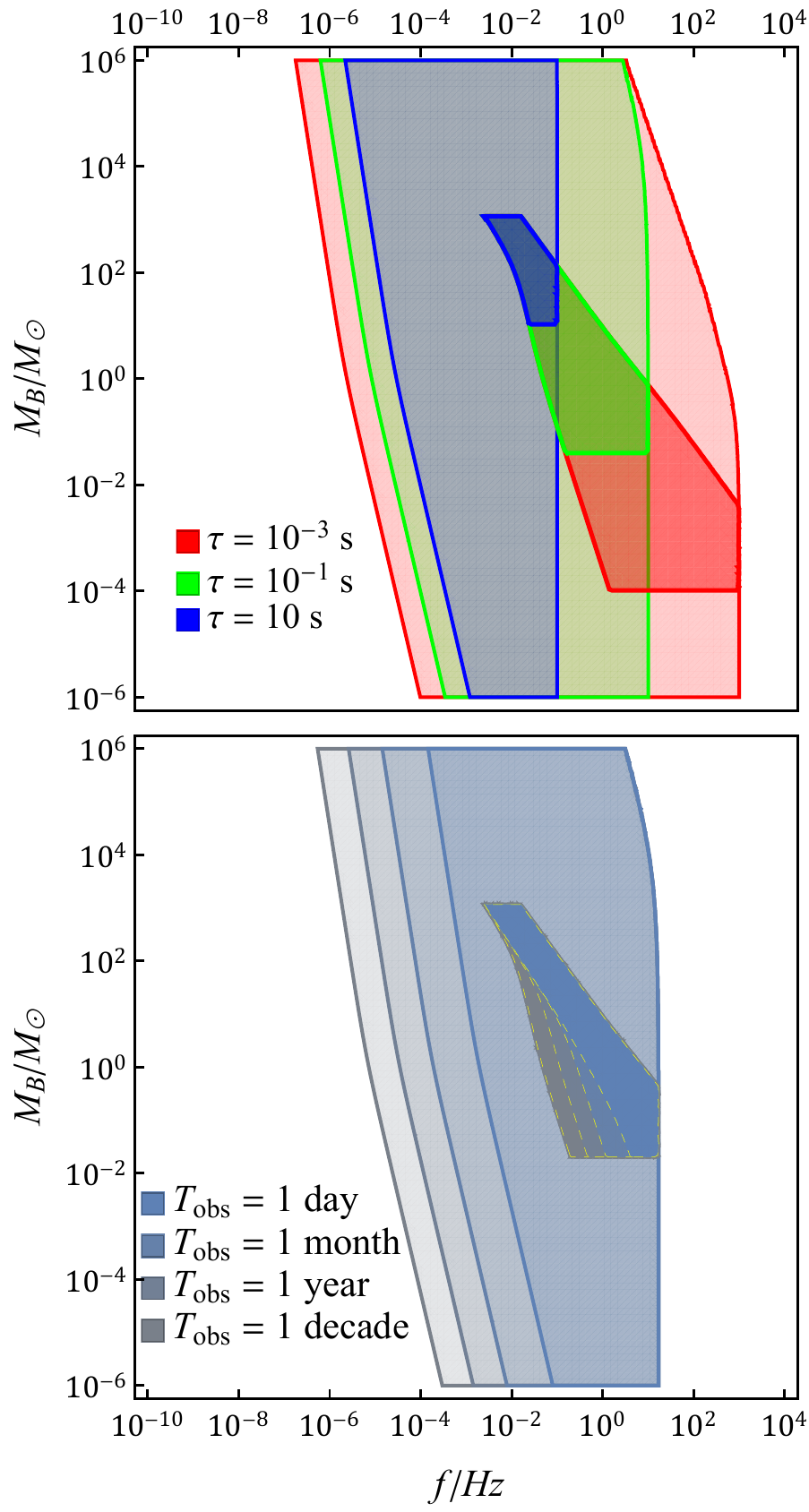}\\
	\caption{The dark shades show the initial orbital frequency $f_\text{ini}$ and BH mass $M_B$ that can possibly leads to detection of the $|322\rangle\to |200\rangle$ resonance within $T_\text{obs}$ of observation. On the upper panel, we pick $T_\text{obs}=10\text{ yrs}$ and change the pulsar rotation period $\tau$. On the lower panel, we pick $\tau=0.1\text{ s}$ and change the maximal observation time. The light opaque shades are for testing GW emission alone as obtained in Fig.~\ref{figfMtauAndfigfMTobs}.}\label{figfloatingfini}
\end{figure}

As a further simplification, we fix the pulsar mass to be $M_P=1.4 M_\odot$. Thus there are five free parameters left, $i.e.$, $\alpha$, $M_B$, $P_{\text{ini}}$, $\tau$ and $T_{\text{obs}}$. We can fix two parameters ($\tau$ and $T_{\text{obs}}$) and integrate out $\alpha$ to obtain a two-dimensional region, out of which GCP detection is excluded. This amounts to a 2-d projection of a 3-d slice of the 5-d region in the parameter space. To observe GCP transitions, $P_\text{ini}$ cannot be too large or too small, namely during the total observation time, $P(t)$ solved from (\ref{GCPevolutionEq}) with an initial value $P(0)=P_\text{ini}$ must at least reach the resonance band. Therefore, the criterion to see the resonance is given by
\begin{equation}
	\text{(C5):  }P_\text{ini}>P_r-\frac{1}{2}\Delta P_r,\text{ and }P(T_\text{obs})<P_r+\frac{1}{2}\Delta P_r~.
\end{equation}
Now combining (C1$-$C5), we can obtain a shaded area on the $f_\text{ini}$-$M_B$ plane, within which GCP resonance detection is possible. In Fig.~\ref{figfloatingfini}, we choose different values of $\tau$ and $T_{\text{obs}}$ and plot this region together with that obtained considering GR effects and timing accuracy in Sect.~\ref{SandA}

As can be seen from Fig.~\ref{figfloatingfini}, the GCP-feasible region lies well-within that of GR tests. Furthermore, the left edge representing the pulsar-timing accuracy frontier is far from the GCP-feasible region. This confirms again that the timing accuracy is adequate for probing the GCP resonance.

\subsection{Sinking orbit: $|211\rangle\to |31\mathrm{-1}\rangle$}

\begin{figure}[h!]
	\centering
	\includegraphics[width=7.5cm]{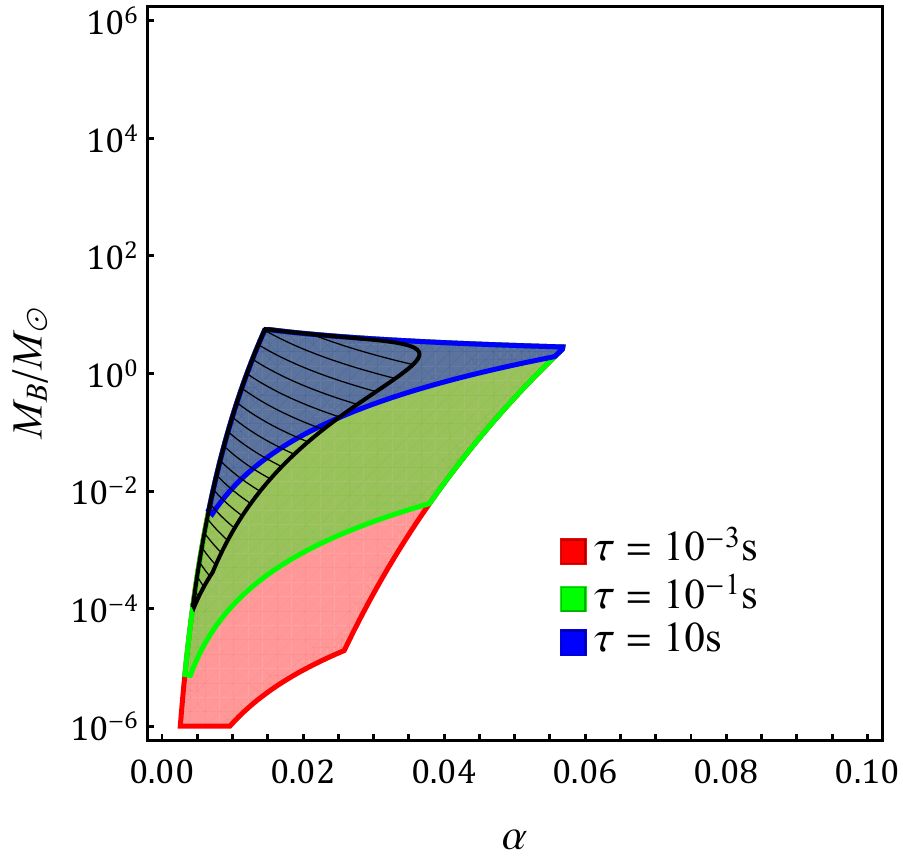}\\
	\caption{The feasible parameter region for $\alpha$, $M_B$ and different pulsar rotation periods $\tau$ in the sinking-orbit resonance $|211\rangle\to |31\mathrm{-}1\rangle$. The other parameters are chosen to be $M_P=1.4 M_\odot$ and $T_{\text{obs}}=10 \text{ yrs}$. The shaded area is where $\Delta t+\Delta t_c> T_{\text{obs}}$ and we cannot observe a whole resonance for this shaded region.}\label{figsinkingalphaMB}
\end{figure}

Now we turn to the analysis of the sinking-orbit resonance $|211\rangle\to |31\mathrm{-1}\rangle$. In this resonance, the boson cloud drains energy from the orbiting pulsar and the resonance can easily become non-adiabatic (also known as the kicked orbit \cite{Baumann:2019ztm}) if $\delta>1$. In contrast, the adiabaticity in the floating orbit case is enhanced. The analysis of a non-adiabatic atomic transition is beyond the scope of this work. Henceforth, we choose to impose a further constraint on the adiabaticity of the resonance,
\begin{equation}
	\text{(C6):  }\delta=\frac{\Delta t_c}{\Delta t}<1~.
\end{equation}
In addition, (C1) is now changed to
\begin{equation}
	\text{(C1'):  }T^{(\text{growth})}_{211}\lesssim 10^6 \text{yrs},~~~T^{(\text{deplete})}_{211}\gtrsim 10^8 \text{yrs}~.
\end{equation}
Combining (C1'), (C6) and (C2$-$C4), we obtain the allowed parameter region for this sinking orbit case (see Fig.~\ref{figsinkingalphaMB}).

As a result, in this sinking-orbit case, the fine structure constant $\alpha$ is smaller than that in the floating-orbit case by approximately an order of magnitude. The BH mass is also limited to $M_B\lesssim \mathcal{O}(10) M_\odot$ by (C6). This is because the cloud energy increases with BH mass. The energy extracted from the pulsar would be too much for its orbital motion to be adiabatic.

\begin{figure}[h!]
	\centering
	\includegraphics[width=7.5cm]{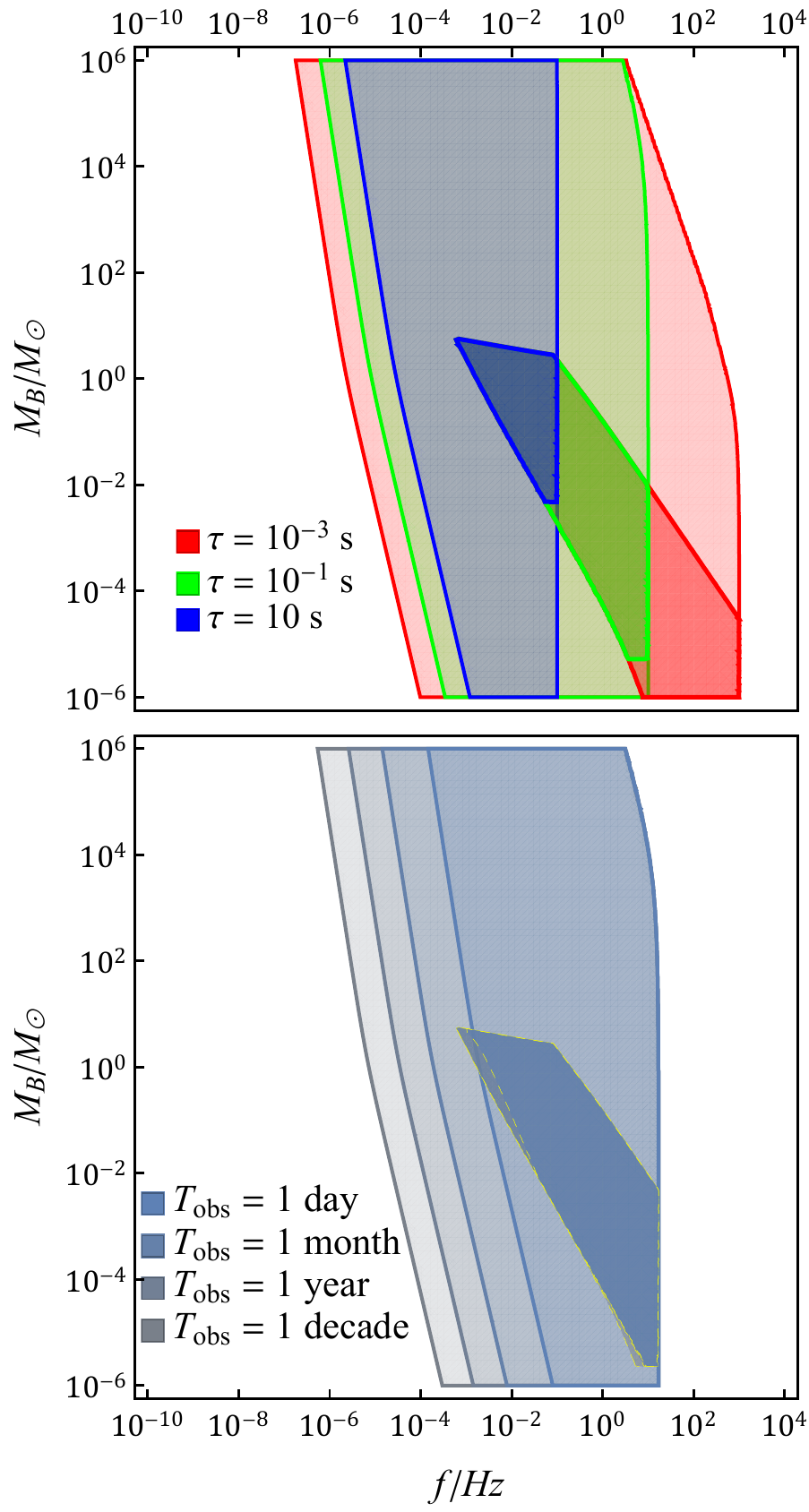}\\
	\caption{The dark shades show the initial orbital frequency $f_\text{ini}$ and BH mass $M_B$ that can possibly leads to detection of the $|211\rangle\to |31\mathrm{-}1\rangle$ resonance within $T_\text{obs}$ of observation. On the upper panel, we pick $T_\text{obs}=10\text{ yrs}$ and change the pulsar rotation period $\tau$. On the lower panel, we pick $\tau=0.1\text{ s}$ and change the maximal observation time. The light opaque shades are for testing GW emission alone as obtained in Fig.~\ref{figfMtauAndfigfMTobs}.}\label{figsinkingfini}
\end{figure}

The estimation of the conditional probability of resonance detection after observing a PSR-BH binary with $M_B$, $M_P$ and $P_\text{ini}$ is analogous to that of the floating-orbit case. The result is plotted in Fig.~\ref{figsinkingfini}.

Again, timing accuracy proves to be not an issue for this sinking resonance case. Note that interestingly, in the lower panel of Fig.~\ref{figsinkingfini} , the detection probability seems to be not sensitive to $T_\text{obs}$, with one day's observation giving almost the same range as a decade's. This does not, however, imply that increasing observation time is ineffective, because this shaded region is merely a projection onto the $f_\text{ini}$-$M_B$ plane. What one can safely deduce is that out of the shaded region, GCP detection is unlikely. Yet for an observed binary lying within the shaded region, whether GCP resonance can be finally detected depends on the unknown probability distribution function of $\alpha$. For a fixed (but unknown $\alpha$), increasing $T_\text{obs}$ surely enhances the detection probability.

\section{Conclusion}\label{conclusion}
In this paper, we have pointed out an alternative way to probe GCP resonance other than the original BH-BH-GW channel. Namely, one can look at the PSR-BH-radio channel, using the same underlying methodology as that of the indirect detection of GWs from the Hulse-Taylor binary. The only difference is that here the GCP detection would be a \textit{direct} one. The Rømer delay measurement is straightforwardly translated into the measurement of the orbital period. The accumulated periastron time shift can deviate from that of the GR prediction in the presence of a GCP resonance. We examined the sensitivity and timing accuracy of this PSR-BH-radio channel and found that for observed $M_B$ and $P_\text{ini}$ lying within the feasible parameter regime, the likelihood of GCP resonance detection is not limited by timing accuracy, but only by the unknown probability distribution of $\alpha$ (or equivalently, the boson mass $\mu$). This suggests that the pulsar-timing accuracy is always enough to capture the resonance phenomenon. 

Roughly speaking, the PSR-BH-radio channel can cover BHs within mass range $10^{-6}M_\odot<M_B<10^3 M_\odot$ and initial period within $10^{-3}\text{ s}<P_\text{ini}<10^3\text{ s}$. This translates into a possible coverage range of boson mass $10^{-14}\text{ eV}<\mu<10^{-6}\text{ eV}$. The shortcoming of this channel is that the current observable range of radio telescopes is mostly limited to our own galaxy. However, the fast developments of radio astronomy in the recent years provides hope for future observation of pulsars outside the galaxy, possibly also PSR-BH binaries.

We note that one very important aspect of GCP is to observe multiple transitions, which is not studied in our present work. The frequency ratio between two sequential resonances is a rational number that is only determined by the transition quantum numbers. Determining these frequency ratios is vital for breaking the parameter degenerates. Therefore, a careful study of observing multiple resonances via the PSR-BH-radio channel is necessary in the future.

We have discussed the observation of GCP from pulsar signals only. If aligned with gravitational wave signals, we should have a better chance of observing GCP transitions. For example, if the gravitational wave detectors can locate the position of the PSR-BH system, that will allow the radio telescopes to point to the right direction to search for this PSR-BH system and the associated transition events.

Another interesting aspect is that allowing the possibility of PBHs also calls for a new mechanism of PSR-BH binary formation. Namely, instead of starting with a binary star system, a pulsar can capture a light PBH, thus forming a binary. Therefore, the detection of GCP resonances with $M_B\lesssim M_\odot$ can serve as an evidence for the existence of PBHs. In this regard, we can consider GCP both as a probe of ultralight bosons and as a smoking gun for PBHs. We leave the study of such a dual picture to future works.

\section*{Acknowledgement} 
We would like to thank Leo Wing-Hong FUNG and Hoang Nhan Luu for helpful discussions.

\bibliographystyle{utphys}
\bibliography{reference}

\end{document}